\documentclass[a4paper,11pt]{article}
\usepackage{pos}
\usepackage{aps_macros}
\usepackage{float}

\newcommand{\npp}{\texttt{nuPyProp}~}
\newcommand{\nss}{\texttt{nuSpaceSim}~}

\title{Monte Carlo simulations of neutrino and charged lepton propagation in the Earth with \npp}
\ShortTitle{Monte Carlo simulations with \npp}

\author*[a]{Sameer Patel}
\author[a]{Mary Hall Reno}

\affiliation[a]{University of Iowa,\\
  Department of Physics and Astronomy, Iowa City, USA}

\forColl{NuSpaceSim} 

\emailAdd{sameer-patel-1@uiowa.edu}
\emailAdd{mary-hall-reno@uiowa.edu}

\abstract{An accurate modeling of neutrino flux attenuation and the distribution of leptons they produce in transit through the Earth is an essential component to determine neutrino flux sensitivities of underground, sub-orbital and space-based detectors. Through neutrino oscillations over cosmic distances, astrophysical neutrino sources are expected to produce nearly equal fluxes of electron, muon and tau neutrinos. Of particular interest are tau neutrinos that interact in the Earth at modest slant depths to produce $\tau$-leptons. Some $\tau$-leptons emerge from the Earth and decay in the atmosphere to produce extensive air showers. Future balloon-borne and satellite-based optical Cherenkov neutrino telescopes will be sensitive to upward air showers from tau neutrino induced $\tau$-lepton decays. We present \texttt{nuPyProp}, a python code that is part of the \nss package. \npp generates look-up tables for exit probabilities and energy distributions for $\nu_\tau\to \tau$ and $\nu_\mu\to \mu$ propagation in the Earth.  This flexible code runs with either stochastic or continuous electromagnetic energy losses for the lepton transit through the Earth. Current neutrino cross section models and energy loss models are included along with templates for user input of other models. Results from \npp are compared with other recent simulation packages for neutrino and charged lepton propagation. Sources of modeling uncertainties are described and quantified.}

\FullConference{37$^{\rm{th}}$ International Cosmic Ray Conference (ICRC 2021)\\
		July 12th -- 23rd, 2021\\
		Online -- Berlin, Germany}

\begin{document}
\maketitle


\section{Introduction}

In the era of multi-messenger astronomy, more focus is on observing the Universe with neutrinos as cosmic messengers. Cosmic neutrinos with energies above a PeV are produced either within astrophysical sources or when ultrahigh energy (UHE) cosmic rays interact with the cosmic background radiation. Observations of these neutrinos are key to understanding cosmic ray acceleration, composition and source evolution. While the production of tau neutrinos is highly suppressed from astrophysical sources, flavor mixing coupled with long baselines allows us to observe all the flavors of neutrinos on the Earth with an equal (flavor) ratio of approximately 1:1:1 \citep{flavor_mixing}. Various neutrino observatories such as IceCube \cite{icecube_sources}, KM3Net \cite{Huang:2020bdg}, ANITA \cite{anita_sources} 
and Baikal \cite{baikal} collect data to find the neutrino emitters in our Universe. The current sky map for detected point sources of neutrino emission is relatively sparse when compared to optical or even high energy photon emitters. This motivates the development and design of new experiments and detectors such as POEMMA \cite{Olinto:2020oky}, IceCube Gen 2 \cite{icecube_gen2} and GRAND \cite{grand}, which have better sensitivities to neutrino fluxes that range to higher neutrino energies, with a goal to determine the population of high energy neutrino emitters in the sky. One method to observe neutrinos makes use of tau neutrinos skimming the Earth, which create upward going extensive air showers (EASs) \cite{fargion} in the Earth's atmosphere upon exit, created by the decay of $\tau$-leptons. This technique uses the Earth as a tau neutrino converter. Sub-orbital missions such as EUSO-SPB 2 \cite{euso_spb2} and space based missions like POEMMA \cite{Olinto:2020oky} will leverage the large surface area of the Earth to detect optical Cherenkov signals from cosmic neutrinos and do complementary observations of UHE cosmic rays using fluorescence telescopes. Hence, determining the neutrino flux sensitivities in these experiments and missions becomes an important component, which requires an end-to-end simulation package for cosmic neutrino induced EASs.

\begin{figure}[H]
       \includegraphics[width=.4\textwidth]{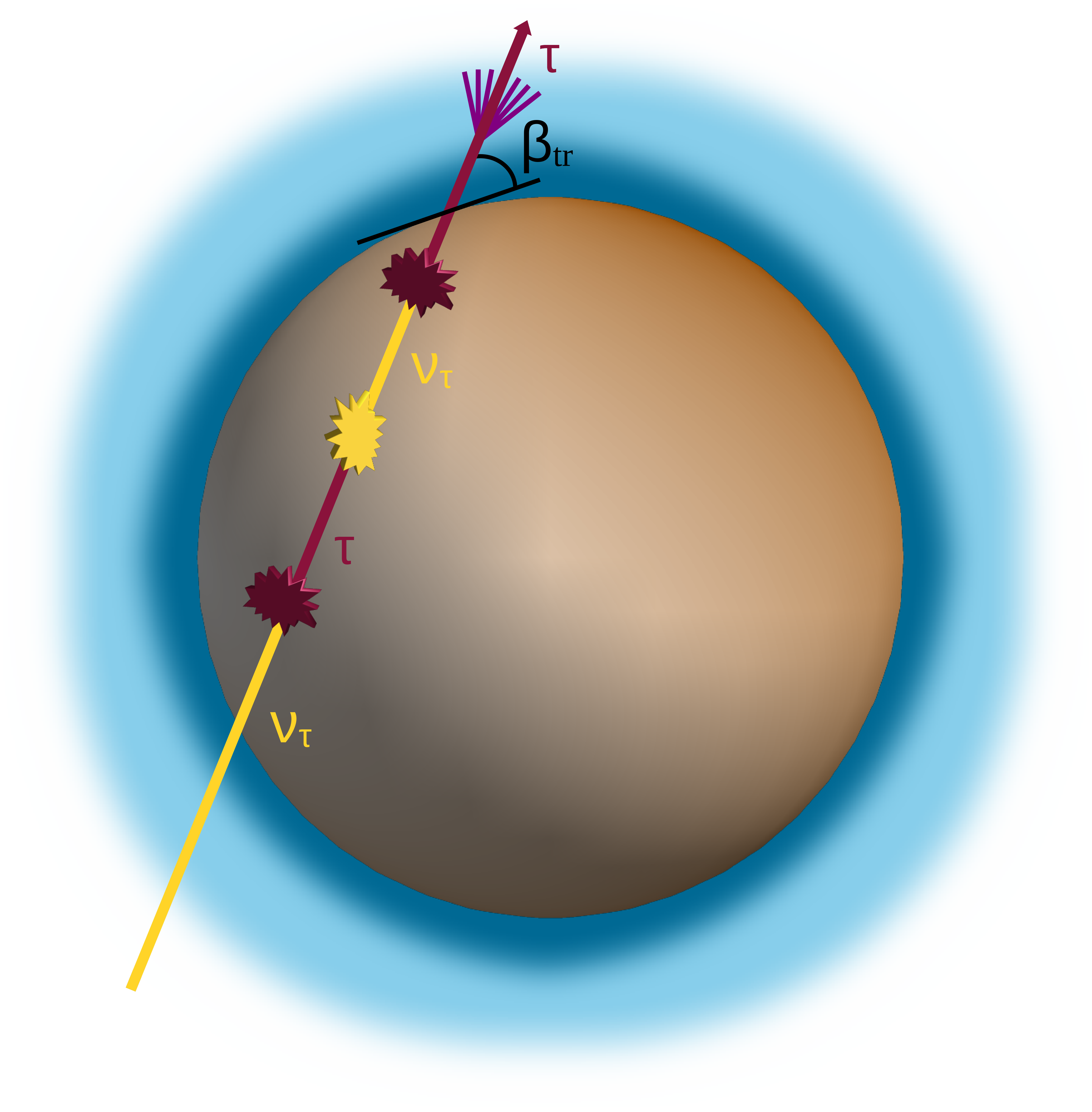}
       \includegraphics[width=.4\textwidth]{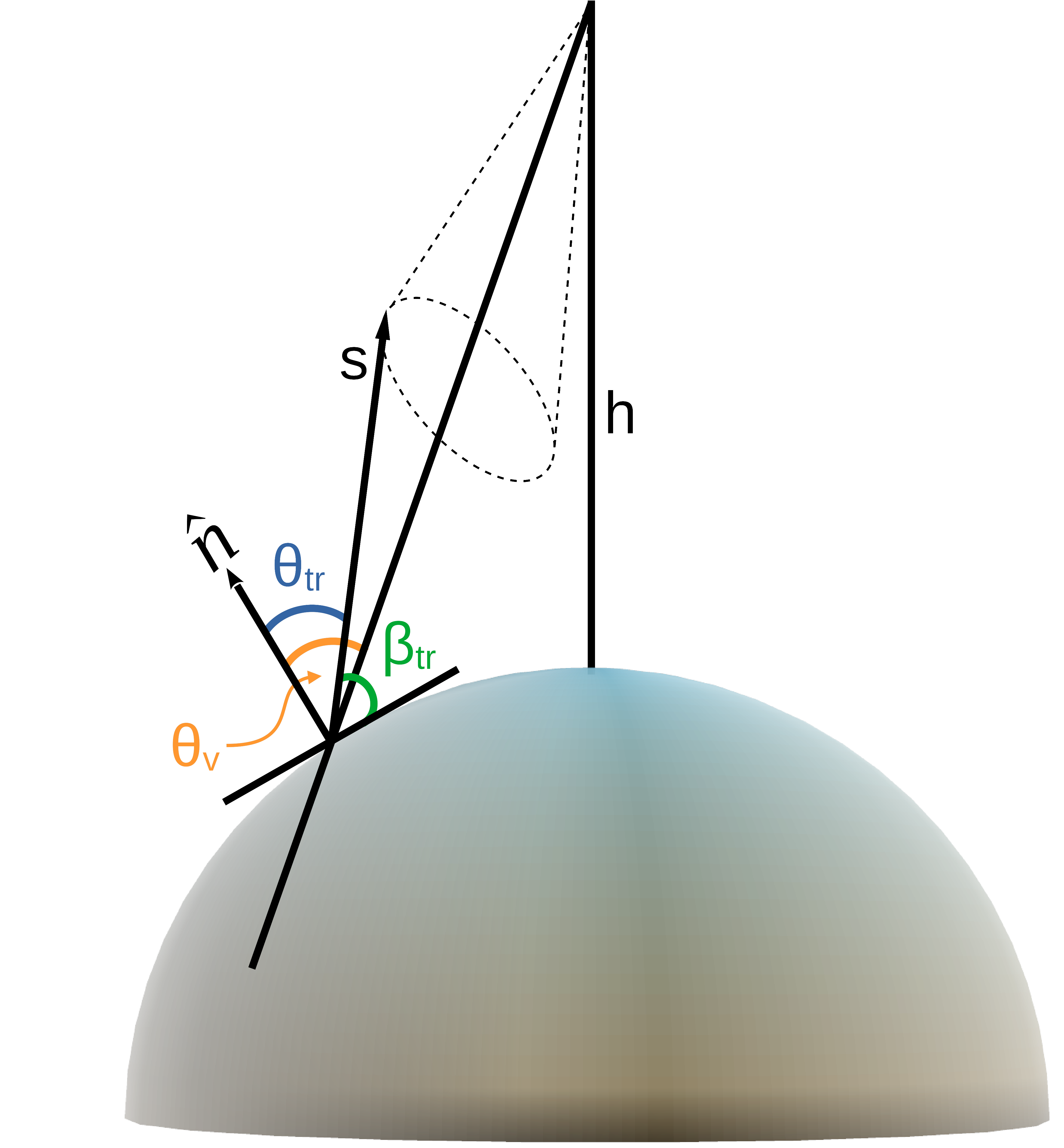}       
       \caption{
      Left: The propagation of $\nu_\tau \to \tau$ inside the Earth. Right: The geometry for observing a $\tau$-lepton at an altitude of $h$ above the Earth's surface. 
       }
       \label{fig:geo}
     \end{figure}

Using a multimessenger approach and rapid slew speeds, POEMMA will be able to follow up transient ToO events \cite{Venters:2019xwi}. This can be done by following the geometry of Earth-skimming tau neutrinos in fig. \ref{fig:geo}. The propagation and energy loss of neutrinos and $\tau$-leptons inside the Earth is discussed in section \ref{sec:prop}. Once a $\tau$-lepton exits the surface of the Earth, the tau observation probability can be written in terms of the exit probability p$_{\text{exit}}$, detection probability p$_{\text{det}}$, and the decay probability p$_{\text{decay}}$ for an infinitesimal path length $ds$ \cite{Reno:2019jtr}:
\begin{equation}
    \text{P}_{\text{obs}} = \int \text{p}_{\text{exit}} (E_{\tau}|E_{\nu_{\tau}},\beta_{tr}) \times \left[\int ds'~\text{p}_{\text{decay}}(s')~\text{p}_{\text{det}}(E_{\nu_{\tau}},\theta_{\text{v}},\beta_{tr},s')\right] dE_{\tau},
\end{equation}
where $\beta_{tr}$ denotes the Earth emergence angle. Here, p$_{\rm decay}$ relates to the decay of the $\tau$-lepton in the Earth's atmosphere as a function of altitude, and p$_{\rm det}$ essentially determines how much of the Cherenkov signal would be effectively observed at the detector [p\_det needs revision?]

The exit probability is an important part of obtaining the observation probability and it is independent of the detector, which allows us to develop a mission/detector-independent simulation package and is one of the main motivations for the development of \texttt{nuPyProp}. We focus on $\tau$-leptons because they are responsible for creating energetic EASs. We extend our results to include muons, whose signals can enhance the detection sensitivity towards lower energy neutrinos \cite{Cummings:2020ycz}.

     
\section{Code Overview}
     
The \npp code is a Monte Carlo package used to propagate neutrinos and leptons inside the Earth and determine the outgoing exit probability as well as the outgoing lepton energy distribution. It is designed to be a highly flexible and modular code. It is a part of the \nss package \cite{nuspacesim} designed to simulate radio and optical signals from EAS that are induced by astrophysical neutrinos. The code package is also available to be used independently. It is written mostly in Python, with the exception of the propagation module, which is written in Fortran90 (wrapped using F2PY\footnote{\href{https://numpy.org/doc/stable/f2py/}{https://numpy.org/doc/stable/f2py/}}) in conjunction with OpenMPI\footnote{\href{https://www.open-mpi.org/}{https://www.open-mpi.org/}} that leverages the use of multi-core CPUs. This retains the high readability of the Pythonic substructure for use in defining models and data manipulations. Table \ref{tab:models} shows the various different theoretical models we implement. We emphasize that user-defined custom models can be easily included by following the templates provided in the code. With the flexibility of models and propagation parameters, \npp can be used for any underground, sub-orbital or space-based detector. While \npp was originally designed to simulate $\nu_\tau \to \tau$-leptons only, the modularity of the code allows simulations of Earth-skimming $\nu_\mu \to \mu$ as well. The code is currently available on GitHub\footnote{\href{https://github.com/NuSpaceSim/nupyprop}{https://github.com/NuSpaceSim/nupyprop}}, and the resulting exit probability and energy cumulative distribution function (CDF) lookup tables will also be included as part of the nuSpaceSim package. Moreover, users who want to query the code results for a predetermined set of run parameters will be able to do so at NASA's HEASARC\footnote{\href{https://heasarc.gsfc.nasa.gov/docs/nuSpaceSim/}{https://heasarc.gsfc.nasa.gov/docs/nuSpaceSim/}}. 

\section{Propagation}
\label{sec:prop}

\begin{figure}[H]
       \includegraphics[width=.99\textwidth]{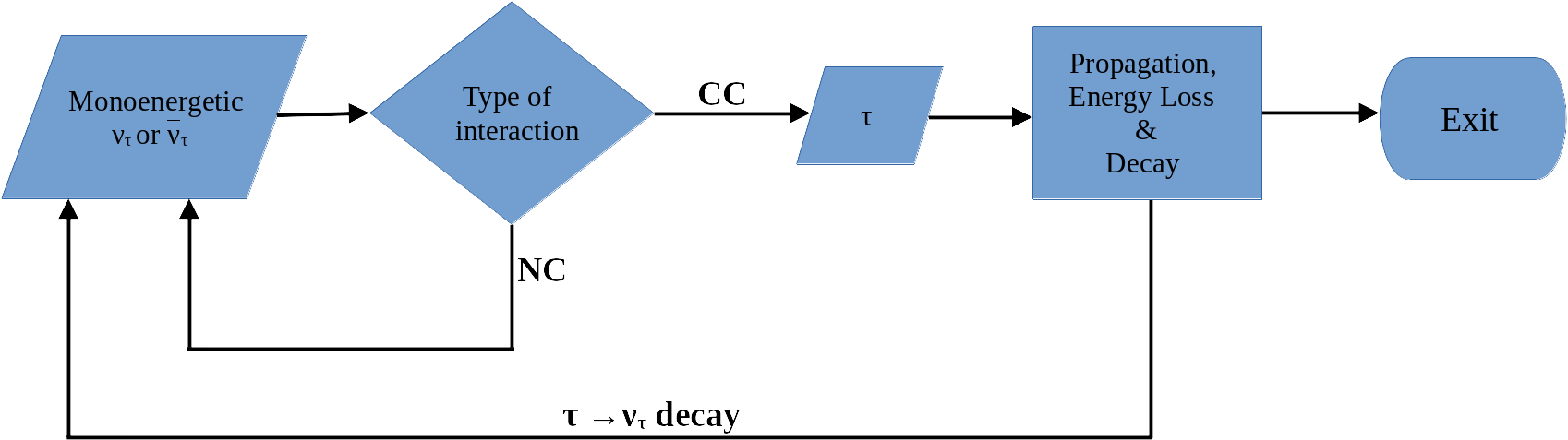}
       \caption{Neutrino and $\tau$-lepton
      propagation flowchart.
       }
       \label{fig:flow}
     \end{figure}
     
As in fig. \ref{fig:flow}, we inject the code with mono-energetic tau or muon neutrinos (or anti-neutrinos) and allow them to propagate, first through a layer of water and then into the inner layers of the Earth with the material densities varying between that of rock and iron. The neutrinos can either convert to same-flavor leptons if there is a charged current interaction or lose energy in case of a neutral current interaction with nucleons. Several neutrino cross section evaluations with different high energy extrapolations are options, as indicated in Table \ref{tab:models}.
After neutrino charged-current interactions, the charged leptons lose energy through either ionization, bremsstrahlung, pair production or photonuclear energy loss. High energy extrapolations of the photonuclear cross section are uncertain, so three energy loss models
\cite{allm,bdhm,ckmt} (ALLM, BDHM, CKMT) that account for different parameterizations of the electromagnetic structure function $F_2$ are included. The Bezrukov and Bugaev (BB) photonuclear energy loss formula was used historically \cite{pn_bb}.
Depending on their initial energies, charged leptons may also decay back into neutrinos (with lower energies), which is commonly known as regeneration.
Regeneration is only important for taus. The charged leptons that make it through the Earth are tracked and their energies recorded. We focus on the energy loss mechanism being stochastic as opposed to a continuous energy loss because the former reflects the more physical interactions that happen inside the Earth. Stochastic energy loss for muons has been known for a while \cite{Lipari:1991ut}. Stochastic energy loss  decreases the charged lepton's range of propagation and reduces its exit probability. In the next section we discuss some of the results from \npp and compare them to known code packages.

\begin{table}[H]
    \centering
\begin{tabular}{|p{0.3\linewidth} | p{0.6\linewidth}|}
    \hline
\textbf{Module} & \textbf{Model/Type} \\ \hline
Earth/Geometry & PREM \cite{prem}, User defined \\ \hline
Neutrino/Anti-Neutrino Cross Section & ALLM \cite{allm,hls_param}, BDHM \cite{bdhm,hls_param}, CTEQ18-NLO \cite{cteq18_nlo, hls_param}, CTW \cite{ctw}, nCTEQ15 \cite{ncteq15,hls_param}, User Defined \\ \hline
Lepton Photonuclear Energy Loss [F$_2$(x,Q$^2$), except BB] & BB \cite{pn_bb,reno_beta}, ALLM \cite{allm,reno_beta}, BDHM \cite{bdhm,reno_beta}, CKMT \cite{ckmt,reno_beta}, User Defined \\ \hline
Electromagnetic Energy Loss Mechanism & Stochastic, Continuous \\ \hline
\end{tabular}
\caption{Earth density, neutrino cross section and electromagnetic energy loss model options in \npp}.
\label{tab:models}
\end{table}

     
\section{Results and Comparisons}
\begin{figure}[!ht]
       \includegraphics[width=.99\textwidth]{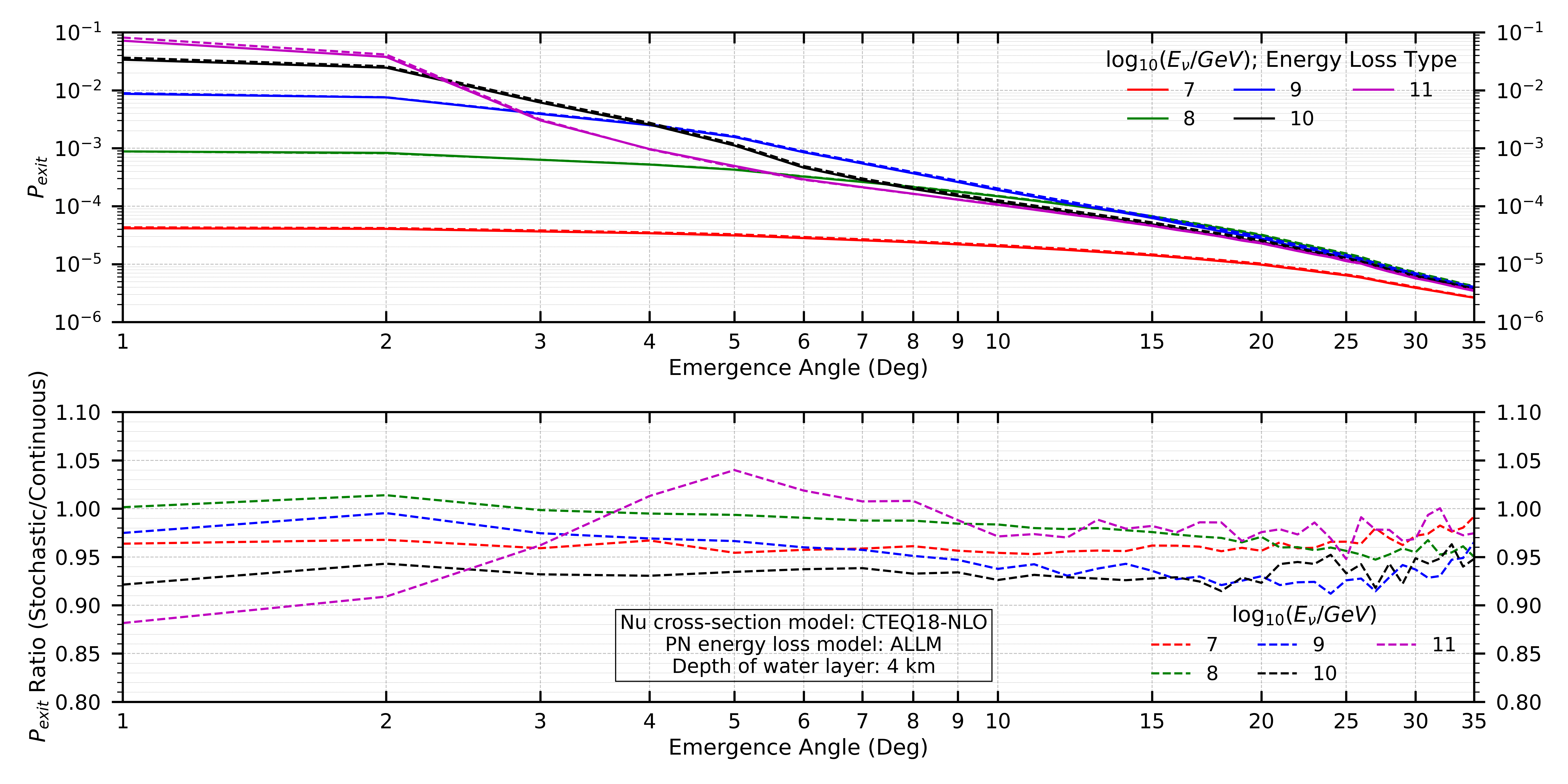}
       \caption{
      Exit probability $P_{\rm exit}$ versus Earth emergence angle $\beta_{\rm tr}$ for stochastic (solid) and continuous (dashed) energy loss modeling ($\nu_\tau \to \tau$)
      (upper) and the ratio of stochastic to continuous evaluation of $P_{\rm exit}$  (lower).
       }
       \label{fig:pexit-sc}
     \end{figure}

\begin{figure}[!ht]
       \includegraphics[width=.99\textwidth]{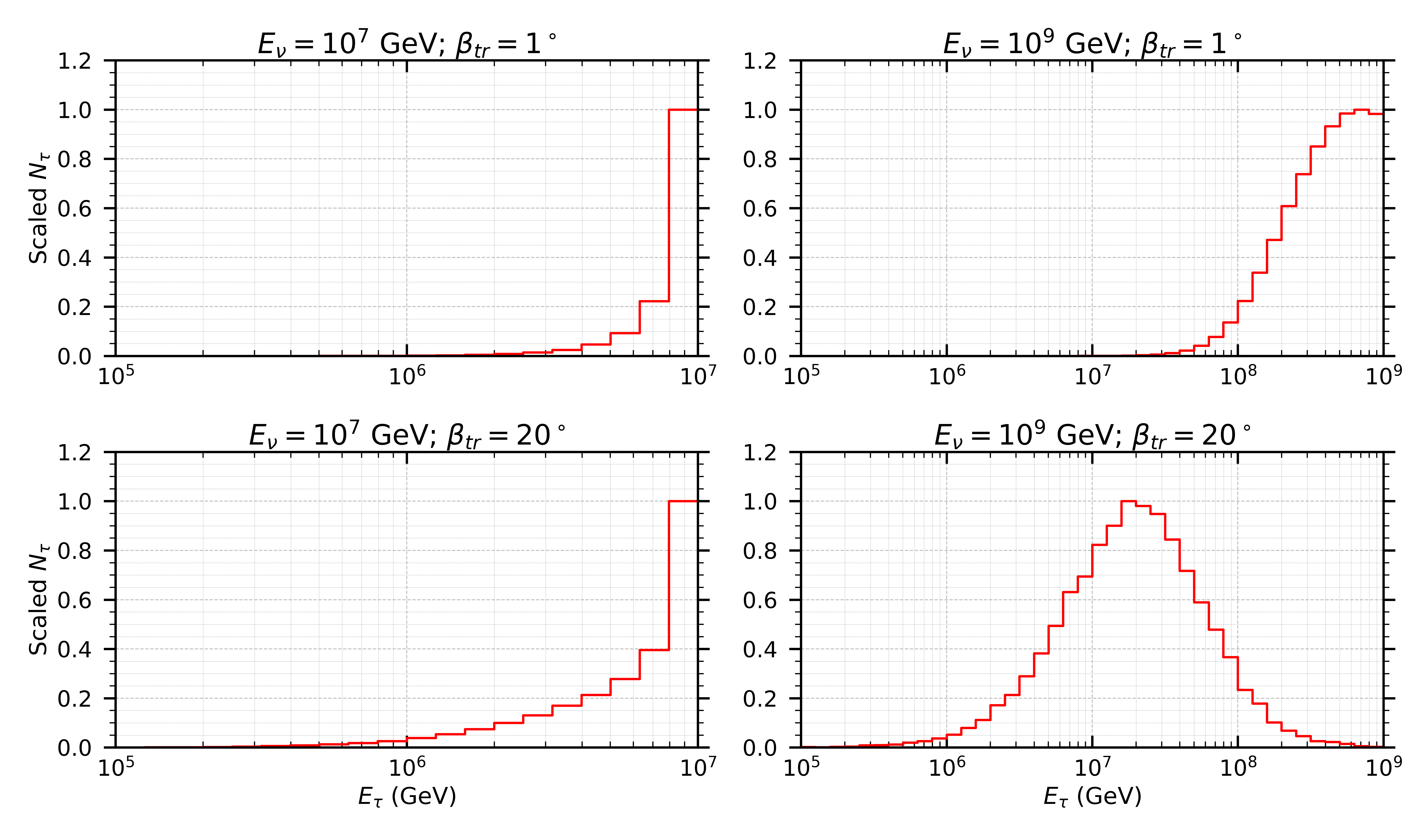}
       \caption{
      Outgoing $\tau$-lepton energy distributions for incident $E_\nu=10^7$ GeV (left) and $E_\nu=10^9$ GeV (right), with $\beta_{\rm tr}=1^\circ$ (upper) and
      $\beta_{\rm tr}=20^\circ$ (lower).
       }
       \label{fig:eout}
     \end{figure}

We show selected results and comparisons here. The exit probability decreases as a function of Earth emergence angle as the neutrino flux is attenuated. We see a decrease of $\sim$10\% in the exit probability  with stochastic energy loss compared to continuous energy loss for $\tau$-leptons, as shown in the $P_{\rm exit}$ ratio in the lower panel of fig. \ref{fig:pexit-sc}. The effect is more prominent at higher neutrino energies and the result is a lateral shift to lower exit probabilities. Fig. \ref{fig:eout} shows the energy distributions of the outgoing $\tau$-leptons for $10^7$ and $10^9$ GeV incoming $\nu_\tau$'s, with Earth emergence angles of $1^\circ$ and $20^\circ$. It is evident here that at larger Earth-emergence angles, because of higher lepton energy losses due to larger column depths traversed, the distribution becomes broader, meaning most of the $\tau$-leptons emerging out of the Earth's surface have been scattered to lower energies compared to the initial neutrino energies.

\begin{figure}[H]
       \includegraphics[width=.99\textwidth]{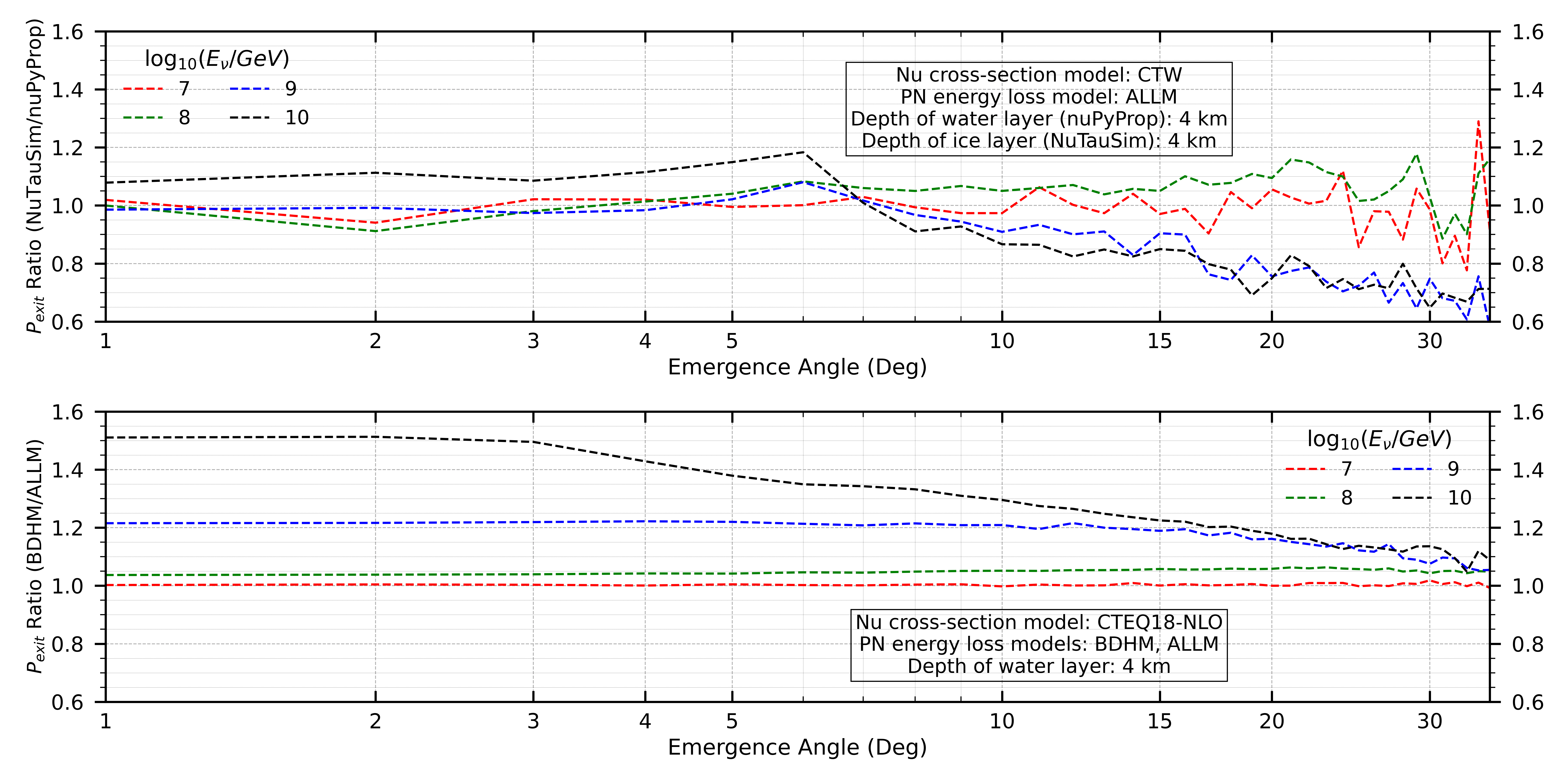}
       \caption{
      Exit probablity ratios for $\nu_\tau\to\tau$  using NuTauSim and nuPyProp (upper) and  BDHM and ALLM photonuclear energy loss models (lower).
       }
       \label{fig:ratios}
\end{figure}
\vspace{-20pt}
\begin{figure}[H]
       \includegraphics[width=.99\textwidth]{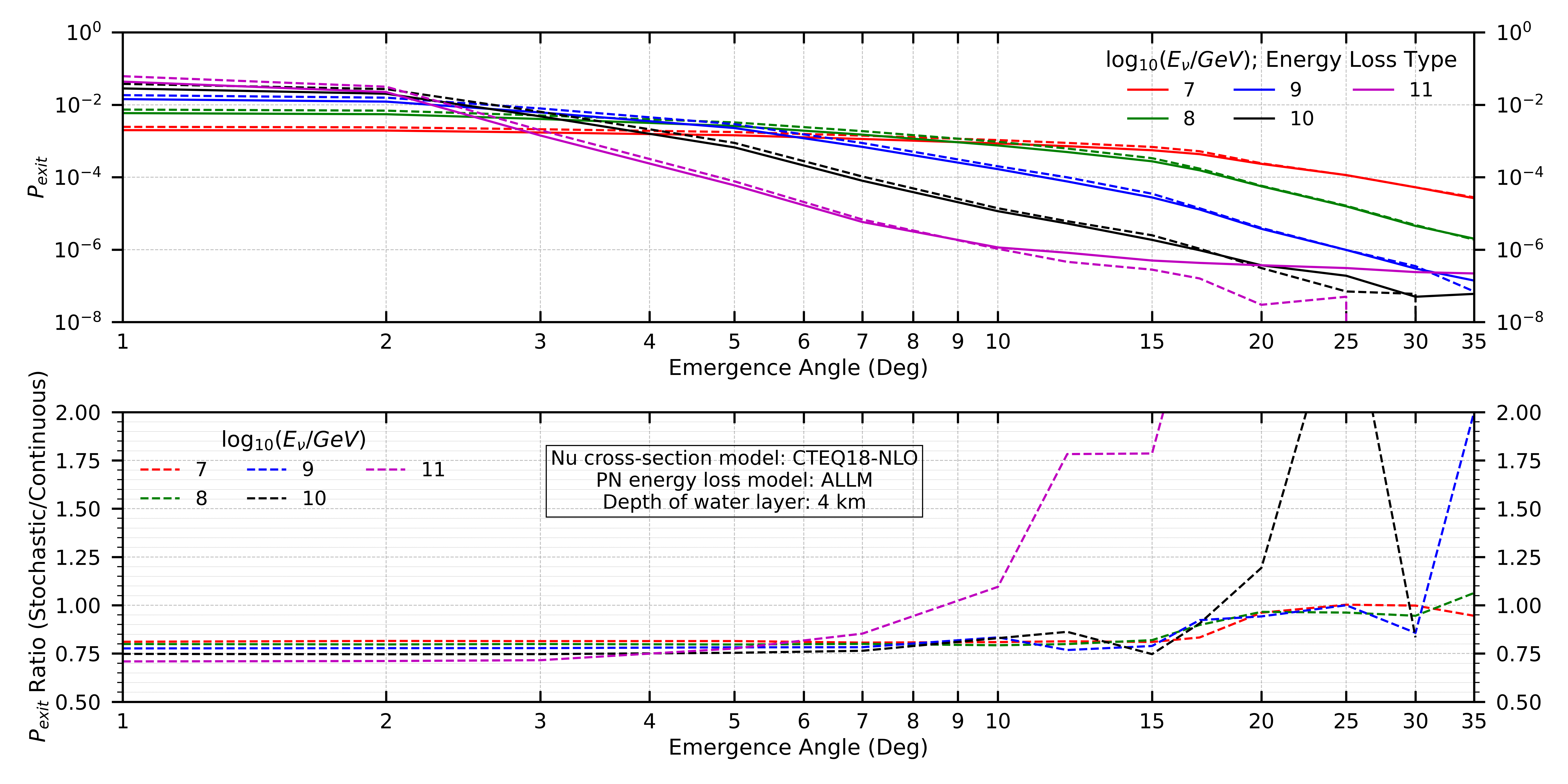}
       \caption{
      Exit probability $P_{\rm exit}$ versus Earth emergence angle $\beta_{\rm tr}$ for stochastic (solid) and continuous (dashed) energy loss modeling ($\nu_\mu \to \mu$)
      (upper) and the ratio of stochastic to continuous evaluation of $P_{\rm exit}$ (lower).}
       \label{fig:pexit-sc-mu}
\end{figure}

Our stochastic results are compared with continuous energy loss results of NuTauSim \cite{nutausim} in the upper panel of fig. \ref{fig:ratios}. We conclude that the differences are based on the different forms of energy loss parameterizations used between the two code packages. Most of the fluctuations at higher energies and angles can be attributed directly to statistical errors from the low number of neutrinos simulated in the NuTauSim package comparison. The lower panel of fig. \ref{fig:ratios} quantifies uncertainties in the photonuclear energy loss models for $\tau$-lepton propagation. It shows the ratio of $P_{\text{exit}}$ with the BDHM energy loss model \cite{bdhm,reno_beta} to that with the ALLM model \cite{allm,reno_beta}. Fig. \ref{fig:pexit-sc-mu} shows the results when we consider the propagation of $\nu_\mu \to \mu$ for stochastic \& continuous energy losses. The result is a $\sim$ 25\% reduction in the exit probability for stochastic losses. At emergence angles $>$10$^\circ$ however, statistical errors start to dominate and warrants a look at higher statistics.

\acknowledgments

This work is supported by NASA grants 80NSSC19K0626 at the University of Maryland, Baltimore County, 80NSSC19K0460 at the Colorado School of Mines, 80NSSC19K0484 at the University of Iowa, and 80NSSC19K0485 at the University of Utah, 80NSSC18K0464 at Lehman College, and under proposal 17-APRA17-0066 at NASA/GSFC and JPL.


\bibliographystyle{apsrev_test}
\bibliography{Bibliography}

\clearpage
\section*{Full Authors List: \Coll\ Collaboration}

%
\scriptsize
\noindent
Yosui Akaike$^1$, 
Luis Anchordoqui$^2$, 
Douglas Bergman$^3$, 
Isaac Buckland$^3$, 
Austin Cummings$^4$,
Johannes Eser$^5$,
Claire Gu\'epin$^6$,
John F. Krizmanic$^{7,8,9}$,
Simon Mackovjak$^{10}$,
Angela Olinto$^5$,
Thomas Paul$^2$, 
Sameer Patel$^{11}$,
Alex Reustle$^{9,12}$,
Andrew Romero-Wolf$^{13}$,
Mary Hall Reno$^{11}$,
Fred Sarazin$^{14}$,
Tonia Venters$^{9}$
Lawrence Wiencke$^{14}$,
and
Stephanie Wissel$^4$ \\

\noindent

$^1$ Waseda Institute for Science and Engineering, Waseda University, Shinjuku, Tokyo, Japan,
$^2$ Department of Physics and Astronomy, Lehman College, City University of New York, New York, New York, 10468 USA,
$^3$ Department of Physics and Astronomy, University of Utah, Salt Lake City, Utah 84112 USA,
$^4$ Department of Physics, Pennsylvania State University, State College, Pennsylvania 16801 USA,
$^5$ Department of Astronomy and Astrophysics University of Chicago, Chicago, Illinois 60637 USA,
$^6$ Department of Astronomy, University of Maryland, College Park, College Park, Maryland 20742 USA,
$^7$ Center for Space Sciences and Technology, University of Maryland, Baltimore County, Baltimore, Maryland 21250 USA,
$^8$ CRESST,
$^9$ NASA/Goddard Space Flight Center, Greenbelt, Maryland 20771 USA,
$^{10}$ Institute of Experimental Physics, Slovak Academy of Sciences, Kosice, Slovakia,
$^{11}$ Department of Physics and Astronomy, University of Iowa, Iowa City, Iowa 52242 USA,
$^{12}$ INNOVIM,
$^{13}$ Jet Propulsion Laboratory, California Institute of Technology, Pasadena, California 91109, USA,
$^{14}$ Department of Physics, Colorado School of Mines, Golden, Colorado 80401 USA

\end{document}